\documentclass[onecollarge,natbib]{svjour2}
\bibpunct{[}{]}{;}{n}{}{,} 
\smartqed  
\usepackage{graphicx,amsmath,amssymb}
\usepackage{color}
\usepackage{enumerate,array}
\definecolor{navyblue}{rgb}{0.3,0.3,1}
\definecolor{purple}{rgb}{0.6,0,0.5}
\usepackage[colorlinks=true, pdfstartview=FitV, linkcolor=purple, citecolor= navyblue, urlcolor=navyblue]{hyperref}

\journalname{Few Body Systems}
\begin{document}

\title{A combined study of the pion's static properties and form factors}

\author{B.~El-Bennich  \and  J. P. B. C.~de Melo \and T.~Frederico}


\institute{Bruno El-Bennich  (\email{bruno.bennich@cruzeirodosul.edu.br})  \at
                Laboratorio de F\'isica Te\'orica e Computacional, Universidade Cruzeiro do Sul, 01506-000, S\~ao Paulo, Brazil
                and Instituto de F\'isica Te\'orica, Universidade Estadual Paulista, 01140-070, S\~ao Paulo, Brazil. \\
                \and Jo\~ao Pacheco B. C. de Melo  \at
                Laboratorio de F\'isica Te\'orica e Computacional, Universidade Cruzeiro do Sul, 01506-000, S\~ao Paulo, Brazil. \\
                \and Tobias Frederico \at
                Departamento de F\'isica, Instituto Tecnol\'ogico de Aeron\'autica, CTA, 12228-900, Sa\~o Jos\'e dos Campos, Brazil.}

\date{Version of \today}

\maketitle

\begin{abstract}

We study consistently the pion's static observables and the elastic and $\gamma^*\gamma \to \pi^0$ transition form factors within a light-front model.
Consistency requires that all calculations are performed within a given model with the same and single adjusted length or mass-scale parameter of the 
associated pion bound-state wave function. Our results agree well with all extent data including recent Belle data on the $\gamma^*\gamma \to \pi^0$ 
form factor at large $q^2$, yet the BaBar data on this transition form factor resists a sensible comparison. We relax the initial constraint on the bound-state 
wave function and show the BaBar data can partially be accommodated. This, however, comes at the cost of a hard elastic form factor not in agreement with 
experiment. Moreover, the pion charge radius is about 40\% smaller than its experimentally determined value. It is argued that a  decreasing charge 
radius produces an ever harder form factor with a bound-state amplitude difficultly reconcilable with soft QCD. We also discuss why vector dominance 
type models for the photon-quark vertex, based on analyticity and crossing symmetry, are unlikely  to reproduce the litigious transition form factor data.

\end{abstract}

\section{Introduction}
\label{intro}

The transition $\gamma^*\gamma \to \pi^0$ has attracted considerable attention with the experimental findings of the BaBar Collaboration~\cite{Aubert:2009mc}.
Indeed, while this new data set appears to agree with earlier experiments on a domain of squared-momentum transfer below $Q^2=-q^2 \lesssim 10$~GeV$^2$
\cite{Behrend:1990sr,Gronberg:1997fj}, the data points at larger $Q^2$ values remarkably exceed the prediction of perturbative QCD (pQCD) in the asymptotic limit
\cite{Farrar:1979aw,Lepage:1980fj}. On the other hand, a most recent measurement by Belle~\cite{Uehara:2012ag} appears to corroborate the pQCD limit.
The process $\gamma^*\gamma \to \pi^0$ is by itself of highly phenomenological and theoretical interest. If the entire domain of experimentally explored squared-momentum 
transfer is to be described within a unique theoretical framework, it must account for the mainly nonperturbative phenomenon of the Abelian anomaly {\em and\/} 
functional behavior of perturbative QCD. This is the challenge for any model employed in a self-consistent way.

Herein we present the results of a consistent treatment of the pion's static features as well as  of the electromagnetic and $\gamma^*\gamma \to \pi^0$ transition
form  factors within one particular model, namely the light-front model introduced in Ref.~\cite{Frederico:1992ye,Frederico:1994dx} and refined in later studies
\cite{deMelo:1997cb,deMelo:1997hh,deMelo:1999gn,deMelo:2003uk,deMelo:2005cy,ElBennich:2008qa,daSilva:2012gf}. This work differs from several recent attempts
\cite{Radyushkin:2009zg,Dorokhov:2010zz,Arriola:2010aq,Kroll:2010bf,Wu:2010zc,Pham:2011jq,Agaev:2010aq,Lichard:2010ap,Gorchtein:2011vf,Melikhov:2012bg,
McKeen:2011aa,Lih:2012yu} to reproduce the large-$Q^2$ BaBar data in spirit and in aim: we do not attempt to build a phenomenological model that yields a 
good fit to the BaBar data. Rather, within the framework of the present model, we insist on a consistent and simultaneous treatment of {\em all\/} extent data, not merely 
the $\gamma^*\gamma\to\pi^0$ transition form factor.

We first calculate the static properties, the pion decay constant and electric charge radius. The pion decay constant, $f_\pi$, serves to adjust the unique parameter
that enters the bound-state wave function, $r_\mathrm{nr}$, and introduces a length (or mass) scale. This immediately fixes the root-mean-square (rms) charge radius of the
pion whose known experimental value, $\langle r_\pi^2 \rangle^{1/2} =  0.672 \pm 0.008$~fm~\cite{Beringer:1900zz}, can be compared with. Next, with this parametrized 
bound state function, we calculate  the leading contribution to the electromagnetic form factor, $F_\pi(Q^2)$, and to the transition form factor, $F_{\gamma\pi}(Q^2)$, in the 
impulse approximation on the light cone. The main freedom in these calculations is, of course, the choice of bound-state wave function or in other words, the
associated Bethe-Salpeter amplitude. 

We devise a wave function which, in interplay with our light-front quark model, reproduces the experimental space-like tail of  $F_\pi(Q^2)$ and thus the asymptotic 
limit of the product $Q^2F_\pi(Q^2) \to 16\pi f_\pi^2\,\alpha_s (Q^2)$ for $Q^2 \to \infty$~\cite{Farrar:1979aw,Lepage:1980fj}. It should be stressed that the decay constant 
and charge radius are insensitive to the asymptotic form of the wave function. This is the reason for the success of quark models or related  contact-interaction models 
in the calculation of the pion's static properties below a mass scale $M^2$~\cite{Bashir:2012fs}. However, probing the pion with $Q^2\gtrsim M^2$ leads to marked deviations 
from experiment  when the contact-interaction model is treated self-consistently with proper regularization \cite{GutierrezGuerrero:2010md,Roberts:2010rn,Roberts:2011wy}. 
In particular, due to the lack of a running mass function, models based on constituent quarks {\em must\/} necessarily introduce a phenomenological bound-state amplitude 
that mimics both soft QCD effects and the asymptotic behavior of the associated pion distribution function, $\phi_\pi(x)$. Therefore, this process also allows us to scrutinize 
and  improve the description of  the pion bound-state function in relativistic light-cone models.

To summarize, we first describe the light-front formulation of the $\gamma^*\gamma \to \pi^0$ and introduce the parametrization of the pion bound-state wave function 
in Sec.~\ref{two}. The values we obtain from this parametrization for $F_{\gamma\pi}(Q^2)$ and $F_\pi(Q^2)$ as well as the pion charge radius and decay constant are 
discussed in Sec.~\ref{three}. We resume in Sec.~\ref{four} where we relax the condition on the length scale parameter and observe the implication on the functional 
behavior of $F_{\gamma\pi}(Q^2)$, in particular whether the asymptotic prediction of perturbative QCD is still verified. We also check the values obtained in this case for 
$\langle r_\pi^2 \rangle^{1/2}$ and $f_\pi$ and modifications occurring in $F_\pi(Q^2)$. In Sec.~\ref{five} we wrap up with a qualitative discussion of the $\gamma^*\gamma \to \pi^0$ 
form factor and its possible evolution upon the inclusion of vector resonances in the deep space-like region.

\section{Pion form factors on the light-front}
\label{two}

\subsection{Preliminaries}

The geometry employed, for example by the BaBar Collaboration, to produce the $\pi^0$ consists of an untagged almost real ($k_2^2 \approx 0$)
photon scattered at small angle from the collision axis and a tagged electron emitting a highly off-shell photon with space-like momentum $k_1$.
The two photons eventually create the neutral pion via the Abelian anomaly.

The relativistic approach to the wave function based on constituents quarks is possible due to the absence of pair-creation processes on the light front.
This salient feature arises from the particular choice of the light-front coordinates~\cite{Dashen:1966zz,Fubini:1966381,Frankfurt:1977vc}, defined by
$x^+ = x^0+x^3$, in which the center-of-mass is readily separated~\cite{Terentev:1976jk}. In essence, the impulse approximation depicted in Fig.~\ref{diagram}
 involves a bound state and an internal loop momentum. The integration over the loop momentum is first performed in the convergent light-front
energy, $k^-$, after which the pion's wave function emerges naturally within the three-dimensional remainder of the integral and depends on 
the $+$ and $\perp$ component of the $\bar qq$ relative momentum~\cite{Frederico:1992ye,Frederico:1994dx,deMelo:1997cb,deMelo:1997hh,Chung:1988my}. 
We stress that in defining a model for this light-front wave function, the convergence of the $k^-$ integration is crucial: it is the case for the good component 
of the current~\cite{Dashen:1966zz,Fubini:1966381} and also for the diagram that describes the weak decay constant.

The aim of this paper is to study models of light-front wave functions in computations of the electromagnetic and transition form factors of the neutral
pion with simultaneous verification of the related weak decay constant and charge radius. We make use of two distinct model wave function:

\begin{enumerate}[i)]
\item a Gaussian model \cite{Chung:1988mu,Chung:1991st};
\item a hydrogen-type wave function \cite{deMelo:1996zn}.
\end{enumerate}
Both wave functions feature a length or mass scale parameter, $r_{\mathrm{nr}}\/$, whose value is of the order of  $\Lambda_{\mathrm{QCD}}$.
We fix this scale with a fit to the pion decay constant, all other observables follow from this parametrization without exception. We compare the results for the form
factors obtained with these two wave functions in Sec.~\ref{three}.

\subsection{Formulation of light-front amplitudes and form factors}

\begin{figure}[t]
\centering
\includegraphics[scale=0.7]{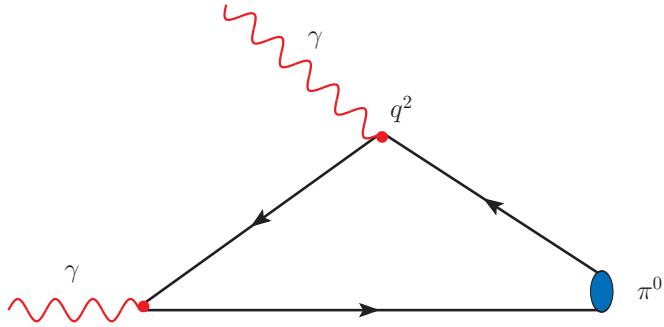}
\caption{The leading diagram corresponding to the impulse approximation of the process $\gamma^* \gamma\to \pi^0$. The lines with arrows denote
constituent light quarks; the solid blobs represent bare photon-quark vertices and the filled oval depicts the $\bar qq$ bound-state wave function of the pion.}
\label{diagram}
\end{figure}

In the following, we formulate the $\gamma^*\gamma \to \pi^0$ transition form factor in the light-front approach. The coupling of the pseudoscalar $\bar qq$
pair to the pion is expressed through the following effective Lagrangian~\cite{Frederico:1992ye,Frederico:1994dx},
\begin{equation}
  \mathcal{L}^{\textrm{int}}_{\pi q}  = -\imath \, \frac{M}{f_\pi} \vec\pi \cdot \bar q\, \gamma^5  \vec\tau\, q \  ,
\end{equation}
where $M$ is the constituent-quark mass, $f_\pi = 92.4$~MeV is the weak decay constant of the pion, $\vec\pi$ and $q$ are respectively the pion field and quark
wave functions. The units are chosen such that $\hbar = c =1$. This coupling can be thought of as the leading term of the full pseudoscalar Bethe-Salpeter amplitude.

The amplitude for the electromagnetic process $\gamma^* \gamma \rightarrow \pi^0$ in the impulse approximation is diagrammatically represented in Fig.~\ref{diagram}
and given by the tensor, $T^{\mu \nu}$, which contains two contributions due to bosonic symmetrization of the amplitude:
\begin{equation}
T_{\mu \nu}(k_1,k_2) = t_{\mu \nu} (k_1,k_2)+ t_{\mu \nu} (k_2,k_1) \ .
\label{tensortot}
\end{equation}
The tensor $t_{\mu \nu} (k_1,k_2)$ is obtained after taking the traces in spinor and flavor space~\cite{Itzykson:1980rh}:
\begin{equation}
  t_{\mu \nu}=\frac{4}{3} \frac{M^2}{f_{\pi}}  e^2_0 N_c \, \epsilon_{\mu \nu \alpha \beta}\,  k_{1}^\alpha k_2^\beta\, I(k_{1}^2) \ ,
\label{tensor}
\end{equation}
where $I(k_1^2)$ is the scalar loop integral,
\begin{equation}
   I(k_1^2)=\int \frac{d^4k}{(2 \pi)^4}  \frac{1}{((k_2 - k)^2 - M^2+\imath \epsilon)} \frac{1}{(k^2 - M^2+\imath \epsilon) ((k_\pi - k)^2 - M^2+\imath \epsilon) } \ .
\label{integral}
\end{equation}
Here, $N_c$ is the number of colors, $k_\pi=k_1+k_2$ is the $\pi^0$ momentum, $k_1\equiv q$ is the space-like momentum transfer, $k_2^2 =0$ is the 
on-shell photon and $e_0$ and $M$ are the unit charge and constituent quark mass, respectively. The factor $1/3$ in Eq.~\eqref{tensor} stems from the flavor trace.

After transformation to light-front variables, $\vec k_{\perp}$, $k^+=k^0+k^3$ and $k^-=k^0-k^3$, we first integrate over $k^-$, the light-front energy.
The reference frame is chosen such that $q^+=q^-=0$ and momentum transfer is transversal, $q_\perp$, which is always possible due to the
space-like character of $q$. In this coordinate system, and after factorizing $k^+$,  $k_\pi^+ -k^+$ and $k_2^+ -k^+$ in the denominator,
the integral in Eq.~\eqref{integral} becomes
\begin{eqnarray}
  I( q^2) & = & \frac{d^4k}{2 (2 \pi)^4}  \int\! dk^- dk^+ d^2 k_\perp  \frac{1}{k^+(k_\pi^+-k^+) (k_2^+ - k^+) \left ( k^{-}- \frac{k_\perp^2 + M^2-\imath \epsilon}{k^+}\right ) }
  \\ \nonumber
           & \times & \frac{1} {\left ( k^-_2 -k^- \frac{(\vec{k}_2-\vec{k})_\perp^2 + M^2-\imath \epsilon}{(k_2^+-k^+)}\right )
           \left (k^{-}_\pi -k^- \frac{(\vec{k}_\pi-\vec{k})_\perp^2 + M^2-\imath \epsilon}{(k_\pi^+-k^+)} \right  )}  \ ,
\label{nullplane}
\end{eqnarray}
where $k_\pi^+=k_2^+$ and $k^-_\pi=k_2^-$ in this reference frame.

The integration in $k^-$, performed using Cauchy's theorem, is convergent. The position of the poles in the $k^-$ complex plane depends on the value of $k^+$.
Since $k_\pi^+ =k_2^+$, this integral is non-vanishing in only one region of $k^+$. The set $k^+ < 0$ and $k^+ > k_\pi^+$ does not contribute to the integral
because the three poles in there have imaginary parts with same signs. The only contribution comes from the region where the $+$ component of the quark
momentum cannot exceed the pion momentum: $0< k^+ < k_\pi^+$. This corresponds to an on-shell quark. The result is
\begin{eqnarray}
\hspace*{-3mm}
  I(q^2) & = & \frac{-\imath}{2 (2 \pi)^3} \int\!  \frac{dx\, d^2 k_\perp}{x(1-x)^2 }
                          \frac{1}{\left (k^-_2 k_2^+ - \frac{k_\perp^2+M^2}{x} -\frac{(\vec{k}_2 - \vec{k})_\perp^2 + M^2}{1 -x}\right )\!\!
                          \left (k^-_\pi k_\pi^+ - \frac{k_\perp^2+M^2}{x} - \frac{(\vec{k}_\pi - \vec{k})_\perp^2 + M^2}{1 -x} \right)}\,   ,
\label{inte2}
\end{eqnarray}
with the momentum fraction, $x=k^+/k^+_\pi$, and $0 < x < 1$.

Introducing in Eq.~\eqref{inte2} the relative transverse $\bar qq$ momentum, as defined in Ref.~\cite{Terentev:1976jk},
\begin{equation}
   \vec{K}_\perp=(1-x) \vec{k}_\perp-x (\vec{k}_\pi-\vec{k})_\perp \ ,
\end{equation}
we have
\begin{equation}
   I(q^2)=\frac{\imath}{2 (2\pi)^3}  \int \frac{dx\, d^2 K_\perp}{x(1-x)}  \frac{1}{((\vec{K}-x \vec{q})^2_\perp+M^2)(m_\pi^2-M_0^2)} \ ,
\label{inte3}
\end{equation}
where $m_\pi$ is the pion mass. The free-mass operator for $\bar qq$ pair is written in terms of the momentum fraction, $x$, and the relative perpendicular momentum as:
\begin{equation}
  M_0^2(K_\perp^2,x) =\frac{K_\perp^2+M^2}{x(1-x)} \ .
 \label{freemass}
\end{equation}

The matrix element of the neutral pion decay driven by the Abelian anomaly is given by one unique $CPT$-invariant Lorentz structure. When one of the photons is
off-shell the same matrix element describes the transition amplitude $\gamma^*\gamma \to \pi^0$,
\begin{equation}
  \langle \pi^0(k_\pi) | J_\mu^{\pi^0} |  \gamma(k_2) \rangle =e^2\, \epsilon_{\mu \nu \alpha \beta}\,  \epsilon^\nu_\gamma q^\alpha k_{\pi}^\beta F_{\gamma\pi^0}(q^2) \ ,
\end{equation}
where  $\epsilon^\nu_\gamma$ is the polarization of the real photon. Using Eqs.~\eqref{tensortot}, \eqref{tensor} and \eqref{inte3}, the  transition form factor is given in our model by,
\begin{equation}
  F_{\gamma\pi^0}(q^2)  =  \frac{N_c}{6 \pi^3}\frac{M^2}{f_\pi}  \int \frac{dx\, d^2 K_\perp}{x(1-x)} \frac{1}{((\vec{K}-x \vec{q})^2_\perp+M^2)(m_\pi^2-M_0^2)} \ .
\label{formfac1}
\end{equation}
In the soft pion limit, following Ref.~\cite{Itzykson:1980rh}, the form factor becomes
\begin{equation}
   F_{\gamma\pi^0}(0)=\frac{1}{4 \pi^2 f_\pi}
\end{equation}
and the neutral pion decay width is given by~\cite{Itzykson:1980rh},
\begin{equation}
   \Gamma_\pi^0=\frac{\alpha^2 m_\pi^3 \pi}{4} F^2_{\gamma\pi^0} (0) \ ,
\end{equation}
where $\alpha$ is the fine structure constant. A charge radius, $\sqrt{\langle r^2_{\pi^0}\rangle}$, can be defined via,
\begin{equation}
   r_{\pi^0}^2 =  6\, \frac{d F_{\gamma\pi^0}(q^2)}{dq^2}\Big |_{q^2=0} \ ,
\label{neutralradius}
\end{equation}
which in the soft-pion limit behaves as~\cite{Ametller:1983ec},
\begin{equation}
 \sqrt{ \langle r^2_{\pi^0} \rangle} =\frac{1}{\sqrt{2} M}\ .
\end{equation}

To complete this section on form factors, we recall the long-standing result on the light front for the elastic pion form factor~\cite{Frederico:1992ye}. Its expression is
likewise obtained following the steps of Eqs.~\eqref{integral} to \eqref{inte3} with the definition of the elastic form factor ($q=p'-p$):
\begin{equation}
 T_\mu(p,p') =  \langle \pi^+ (p') | J^{\pi^+}_\mu | \pi^+(p) \rangle =  e\, F_\pi(q^2) (p + p')_\mu \ .
\end{equation}
The elastic form factor is thus given by
\begin{equation}
   F_\pi (q^2)  = \frac{2N_c}{(2\pi)^3}\frac{M^2}{f_\pi^2}   \int \frac{dx\, d^2 K_\perp}{x(1-x)} \, 
                                              \frac{M_0^2  \left [1+ \frac{(1-x) \vec{q}_\perp \cdot\, \vec{K}_\perp}{\vec{K}^2_\perp +M^2  }\right ]}{( m_\pi^2-M_0'^2) (m_\pi^2-M_0^2)} \ ,
\label{elasticpion}
\end{equation}
where the free mass operators, $M_0^2(K_\perp^2)$ and $M_0'^2(K_\perp'^2)$, were defined in Eq.~\eqref{freemass} in terms of $x$ and the relative perpendicular
momenta, $\vec{K}_\perp=(1-x)(\vec{p}-\vec{k})_\perp-x\vec{k}_\perp$ and  $\vec{K}'_\perp=\vec{K}_\perp+(1-x)\vec{q}_\perp$. The expression for the rms charge radius,
in the soft-pion limit, 
\begin{equation}
  \sqrt{\langle  r_\pi^2 \rangle } = \frac{\sqrt{3}}{2\pi}\frac{1}{f_\pi}\  ,
\label{Tarrach}
\end{equation}
is know as the Tarrach relation~\cite{Tarrach:1979ta}.

\subsection{Wave function models}

We can identify an asymptotic pion wave function in Eqs.~\eqref{formfac1}  and \eqref{elasticpion} and proceed to model one which reproduces soft QCD at low
momentum transfer and hard perturbative effects for large $q^2$. We follow Ref.~\cite{Frederico:1992ye} where the following replacement is in order,
\begin{equation}
   \frac{1}{ -m_\pi^2+M_0^2}\  \longrightarrow \ \frac{\pi^{\frac{3}{2}} f_\pi}{M \sqrt{M_0 N_c}}\, \Phi_\pi(K^2)  \ ,
 \label{asymptoticwave}
\end{equation}
and the wave-function $\Phi(K^2)$  is normalized to one:
\begin{equation}
   \int d^3 K\,  \Phi^2_\pi(K^2) = 1   \ .
\end{equation}
This is the normalization condition within the framework of Hamiltonian light-front dynamics~\cite{Chung:1988my,Fuda1990265}.
The functional dependence of $K^2$ on $\vec{K}_\perp$ and $x$ originates in the free mass operator, $M_0^2$,
\begin{equation}
  K^2 (\vec K_\perp ; x) = \frac{M_0^2}{4}-M^2 \ .
 \label{funcdepK}
\end{equation}
From Eq.~\eqref{formfac1} and Eqs.~\eqref{asymptoticwave} to {\eqref{funcdepK}, we obtain the integral expression for the neutral pion transition form factor on the light front,
\begin{equation}
   F_{\gamma\pi^0}(q^2)=\frac{\sqrt{N_c} M}{6 \pi^{\frac{3}{2}}}\int\! \frac{dx\,d^2K_\perp}{x(1-x) \sqrt{M_0}}  \frac{\Phi_\pi(K^2)}{(\vec{K} - x \vec{q})^2_\perp+M^2} \ ,
\label{formfac}
\end{equation}
In the limit $Q^2 = -q^2 \rightarrow  \infty$, pQCD predicts the form factor decreases asymptotically as $\sim q^{-2}$ with a limiting value,
$Q^2 F_{\gamma\pi^0}(Q^2)= 2 f_{\pi}$~\cite{Farrar:1979aw,Lepage:1980fj}. 



For the purpose of completeness, we also quote the result for the neutral pion decay, $k_1^2=k_2^2=0$. The amplitude for this process can be formulated
using the same reasoning that led to  Eq.~\eqref{formfac},
\begin{equation}
  F_{\pi^0\to\gamma\gamma} = \frac{2}{3} \frac{\sqrt{N_c} M}{ \pi^{3/2}}  \int\! \frac{dx\,d^2K_\perp}{x(1-x) \sqrt{M_0}}
        \frac{1}{((\vec{K} -2x k_\gamma )_\perp^2 + M^2)} \, \Phi_\pi(K^2)\ ,
\label{pi0decay}
\end{equation}
where $|k_\gamma | =m_\pi/2$ and $x \leq 1/2$.
In the soft-pion limit and for a constant pion vertex, $\Phi_\pi \sim (-m_\pi^2+M_0^2)^{-1}$,
\begin{equation}
    F_{\pi^0\to\gamma\gamma} = F_{\gamma\pi^0}(Q^2=-q^2_\perp=0)
\end{equation}
In practice, the calculation of the neutral pion width using either $F_{\gamma\pi^0}(0)$ in Eq.~\eqref{formfac} or $F_{\pi^0\to\gamma\gamma}$ in Eq.~\eqref{pi0decay}
yield the same numerical result within the light-front approach. This follows  from the observation that the pion mass is much smaller than the inverse of the characteristic 
length or mass scale of the wave function.

\begin{table}[t]
\setlength{\extrarowheight}{3pt}

\caption{The model's  length scale parameter, $r_{\mathrm{nr}}$ [Eqs.~\eqref{gauss} and \eqref{h2o}], as a function of the constituent quark masses
and for $f_{\pi}=92.4$~MeV given by Eq.~\eqref{pi0decay}. The corresponding charge radii are listed next to $r_{\mathrm{nr}}$ for both the neutral and charged pion.}
\begin{center}
\begin{tabular}{|c|c|c|c|c|c|c|c|c|c|c|}
\hline
   \ \ Model & $m_{u,d}$~[GeV]   & $r_\mathrm{nr}$~[fm] & $<r_{\pi}^2>^{\!1/2}$ [fm] & $<r_{\pi^0}^2>^{\!1/2}$ [fm] \\
\hline 
Gaussian  & 0.220  & 0.345 & 0.637  & 0.683  \\ 
                   & 0.330  & 0.472 & 0.655  & 0.552 \\
\hline 
Hydrogen &  0.220 & 0.593 & 0.795   & 0.782  \\
                   &  0.330 & 0.708 & 0.807   & 0.582 \\
\hline
Experiment~\cite{Beringer:1900zz}  &  &  & $0.672 \pm 0.008$ & \\
\hline

\end{tabular}
\end{center}
\label{table1}
\end{table}
\begin{table}[t]
\setlength{\extrarowheight}{3pt}
\caption{The models' length scale parameter, $r_{\mathrm{nr}}$ [Eqs.~\eqref{gauss} and \eqref{h2o}], and corresponding pion charge radii for
increasing values of the pion decay constant $f_\pi$. }
\begin{center}
\begin{tabular}{|c|c|c|c|c|c|c|c|c|c|c|c|} 
\hline
   \ \ Model & $f_{\pi}$~[MeV] & $m_{u,d}$~[GeV]   & $r_\mathrm{nr}$~[fm]  & $<r_{\pi}>^{\!1/2}$ [fm] & $<r_{\pi^0}>^{\!1/2}$ [fm] \\
\hline 
\ \ Gaussian & 92.4  & 0.220 & 0.345 & 0.637  &  0.683  \\  
           & 97.0  & 0.220 & 0.303 &  0.589  &   0.657     \\
           & 110.0 & 0.220 & 0.172 &  0.406  &  0.564      \\  
\hline 
\  \ Hydrogen  &  92.4  & 0.220 & 0.593 &  0.795  &  0.782 \\
           & 97.0   & 0.220 & 0.543 &   0.750  & 0.767 \\ 
           & 110.0  & 0.220 & 0.410 & 0.626 & 0.720 \\
\hline 
\ \ Experiment~\cite{Beringer:1900zz}   &  &  &  & $0.672 \pm 0.008$   &   \\
\hline

\end{tabular}
\end{center}
\label{table2}
\end{table}

As mentioned earlier, both form factors are calculated with two different $\bar qq$ bound-state wave functions: the Gaussian and the hydrogen-atom model.
Both models depend on two parameters, the constituent quark mass, $M$ and the length (or mass) scale, $r_{\mathrm{nr}}\/$, of the wave function. In order 
to test the model's mass dependence, all calculations are done with two constituent-quark masses for which we choose $M= m_{u,d} = 220$~MeV and $330$~MeV. 
This length (or mass) scale is set by fitting the pion decay constant, $f_\pi = 92.4$~MeV~\cite{deMelo:1997cb},
\begin{equation}
   f_\pi =  \frac{\sqrt{N_c}M}{4\pi^{3/2}} \int\frac{dx\, d^2k_\perp}{x(1-x) \sqrt{M_0}} \ \Phi_\pi (k^2)
   \label{decayconstant}
\end{equation}
\begin{figure}[t*]
\centering
\includegraphics*[scale=0.5]{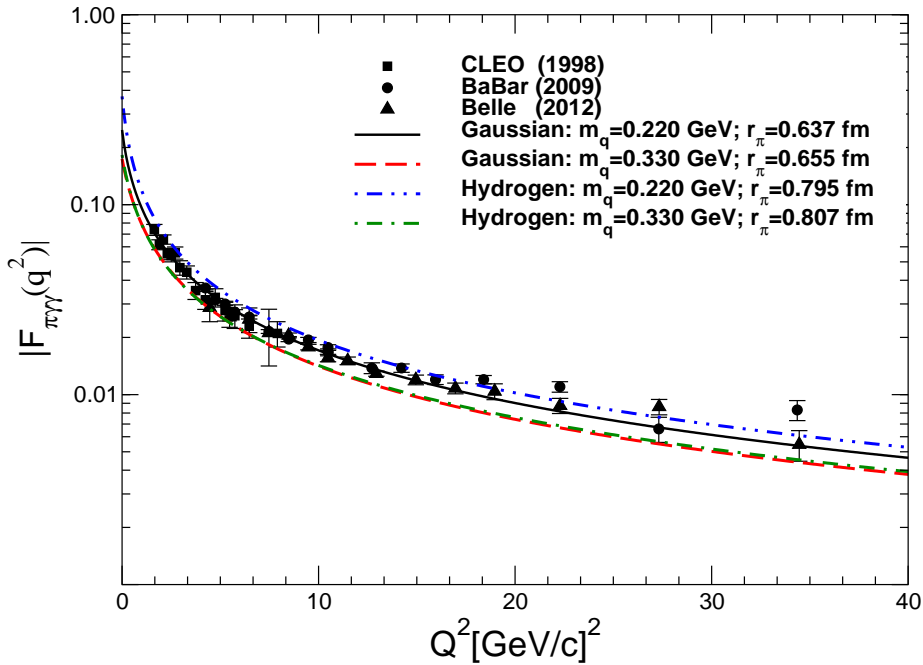}
\caption{The space-like $\gamma^*\gamma\to \pi^0$ transition form factor, $F_{\gamma \pi}(Q^2=-q^2)$, for two different constituent quark masses $m_q=m_{u,d}=M$ and 
$f_\pi = 92.4$~MeV; see Eqs.~\eqref{gauss} and \eqref{h2o} for the two model wave functions and Table~\ref{table1}  for the  corresponding length scale parameter, 
$r_\mathrm{nr}$, and charge radii. Note that the ordinate is scaled logarithmically. Data are from Refs.~\cite{Aubert:2009mc,Gronberg:1997fj,Uehara:2012ag} .}
\label{fig2}
\end{figure}
where the two models for the pion wave function are explicitly given by,
\begin{eqnarray}
     \Phi_\pi (K^2(\vec K_\perp;x))  & = &\ \mathcal{C}_\pi \exp \left [ -\frac{4}{3}\, r_{\mathrm{nr}}^2 K^2 \right ] \ ,  \label{gauss} \\
     \Phi_\pi (K^2(\vec K_\perp;x)) & = &\ \mathcal{C}_\pi  \frac{1}{\left ( r_{\mathrm{nr}}^2+K^2 \right )^2} \ , 
 \label{h2o}
\end{eqnarray}
where $\mathcal{C}_\pi$ is the overall normalization. Once $r_{\mathrm{nr}}$ is fixed, the charge radius \eqref{neutralradius}, the transition form factor \eqref{formfac}, 
and the elastic form factor \eqref{elasticpion} are all calculated  with the same given pion wave function on the light cone. In Section~\ref{three}, we explore 
the model sensitivity to the constituent quark mass with regard to the pion's static observables and elastic and transition form factors. In Section~\ref{four} we study the impact 
of modifying the pion's length scale $r_{nr}$ on the functional behavior of $F_{\gamma\pi^0}(Q^2)$ and the consequences for $F_\pi(Q^2)$ and the rms charge radius.

\begin{figure}[t]
\centering
\includegraphics*[scale=0.5]{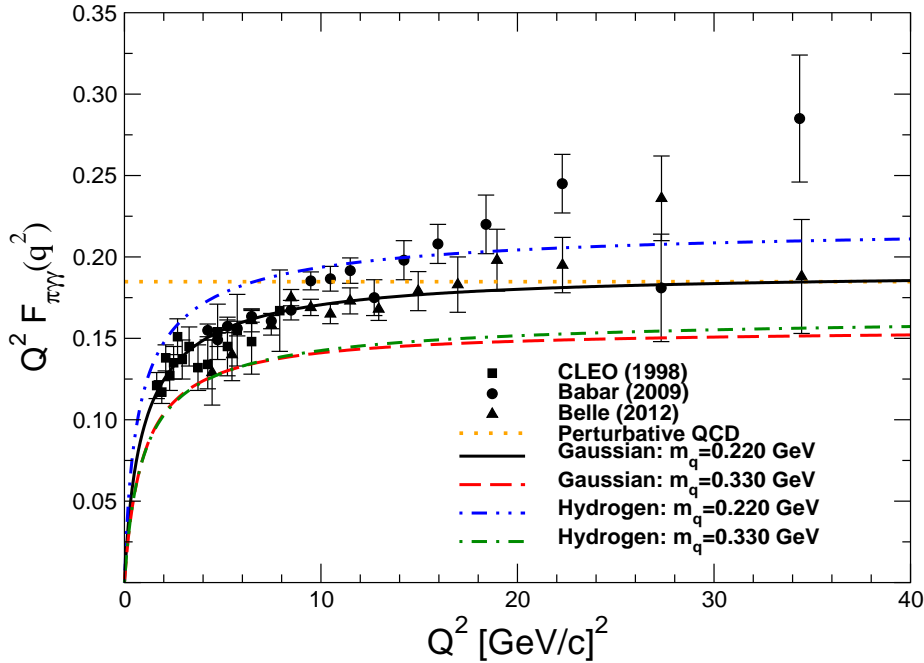}
\caption{The momentum-squared weighted transition form factor, $Q^2 F_{\gamma \pi}(Q^2)$, for two different constituent quark masses $m_q=m_{u,d}=M$ and 
$f_\pi = 92.4$~MeV. The dotted horizontal line is the pQCD prediction~\cite{Lepage:1980fj};  see Eqs.~\eqref{gauss} and \eqref{h2o} for the two model wave functions 
and Table~\ref{table1} for the corresponding length scale parameter, $r_\mathrm{nr}$, and charge radii. Data are from Refs.~\cite{Aubert:2009mc,Gronberg:1997fj,Uehara:2012ag}.
N.B. For $Q^2\to \infty$, $Q^2 F_{\gamma \pi}(Q^2)$ reaches the asymptotic limit $2f_\pi=0.185$~GeV for both models and masses. }
\label{fig3}
\end{figure}

\section{Numerical results and variation of the constituent quark mass}
\label{three}

Numerical results for the transition form factor, $F_{\gamma \pi^0}(Q^2)$, and its weighted description, $Q^2 F_{\gamma \pi^0}(Q^2)$, are presented in Figs.~\ref{fig2} 
and \ref{fig3} for the choice $M =  m_{u,d} = 220$~MeV and $330$~MeV of constituent quark masses. The length scale, $r_{\mathrm{nr}}$, is adjusted  in both cases so as 
to reproduce the pion decay  constant in Eq.~\eqref{decayconstant}. As seen from Table~\ref{table1}, the calculated rms charge radius is in agreement (but too small) with 
the experimental value for the  Gaussian case but about 20\% larger when using the hydrogen model. In the asymptotic limit, pQCD predicts an upper bound 
$\lim_{Q^2\rightarrow\infty}Q^2 F_{\gamma\pi^0}(Q^2)  = 2f_\pi = 0.185$~GeV;  Fig.~\ref{fig3} gives the impression that in all but one case the curves are not bounded by this prediction. 
However, we checked numerically that for large enough values of $Q^2$ all curves converge to $\lim_{Q^2\rightarrow\infty}Q^2 F_{\gamma\pi^0}(Q^2)  = 0.2\ \mathrm{GeV} 
\simeq 2 f_\pi$. We ascribe the difference of about 8\% between this limit and that of pQCD to our modeling of the transversal component of the pion wave function. 

Whilst the hydrogen bound-state model for $m_{u,d}=220$~MeV yields a form factor, 
$Q^2 F_{\gamma \pi^0}(Q^2)$, that comes closest to the BaBar data~\cite{Aubert:2009mc}, it does not provide the logarithmic form these new data points 
seem to imply for $Q^2 > 15$~GeV$^2$. Moreover, for $Q^2 < 10$~GeV$^2$, the hydrogen bound-state model reproduces very poorly older CLEO data. If one does
not take into account BaBar data points at large four-momentum-squared transfer, the Gaussian wave function with $m_{u,d}=220$~MeV provides the best model description 
of all data including Belle's measurements and coincides with the pQCD limit for large $Q^2$.

The elastic form factor  for two constituent quark masses and both models is depicted in Figs.~\ref{fig4} and \ref{fig5}. Here, both the hydrogen and Gaussian models 
with a constituent mass of 220~MeV are in good agreement with data whereas for $m_{u,d}=330$~MeV the form factor appears to be too soft. The hydrogen wave 
function for $m_{u,d}=220$~MeV is particularly successful in reproducing the $1/Q^2$ tail of $F_\pi(Q^2)$. However, as just noted, the resulting charge radius 
$\langle r_\pi^2 \rangle^{1/2} = 0.795$~fm is 20\% larger than the experimental value in this case. On the other hand, this may be a desirable aspect of the model since 
it is crudely given by a quark-antiquark core with a constant dressed quark mass. It is known, however, that dressing the quark-antiquark scattering 
kernel with pion loops introduces an attractive force which results in a decrease of the bound state's mass and charge radius (see, e.g., the discussion 
about pion loop effects in Ref.~\cite{Chang:2009ae}).

We have tested the present light-front model for various constituent masses from 180~MeV to 330~MeV and find the mass range 220--250~MeV to be the most consistent
with a realistic description of the observables discussed herein (for a recent and more detailed discussion, see Ref.~\cite{daSilva:2012gf}). While we acknowledge the mass 
dependence in our calculations, this is hardly limited to the present light-front model~\cite{Melikhov:2001zv}. Indeed, it is a recurrent and unavoidable feature of \textit{all} 
constituent quark models which do not account for dynamical chiral symmetry breaking. The use of a constituent quark mass is justified and yields robust results in calculations 
which involve heavy-quark propagation~\cite{ElBennich:2010ha,ElBennich:2011py,ElBennich:2012tp}, however this is not true for observables involving light quarks
\cite{GutierrezGuerrero:2010md,Roberts:2010rn}. 

Nonetheless, this study aims at a consistent treatment of all form factors, the charge radii and decay constant with the same single parameter set, $r_\mathrm{nr}$ and 
$M=m_{u,d}$. We thus employ in the following $m_{u,d}=220$~MeV which reproduces well $F_{\pi^+}(Q^2)$,  $\langle r_\pi\rangle$  and $f_\pi$ and yields a satisfying 
description of $F_{\gamma \pi}(Q^2)$ below $Q^2 \simeq 15$~GeV$^2$.

\begin{figure}[t]
\centering
\includegraphics*[scale=0.5]{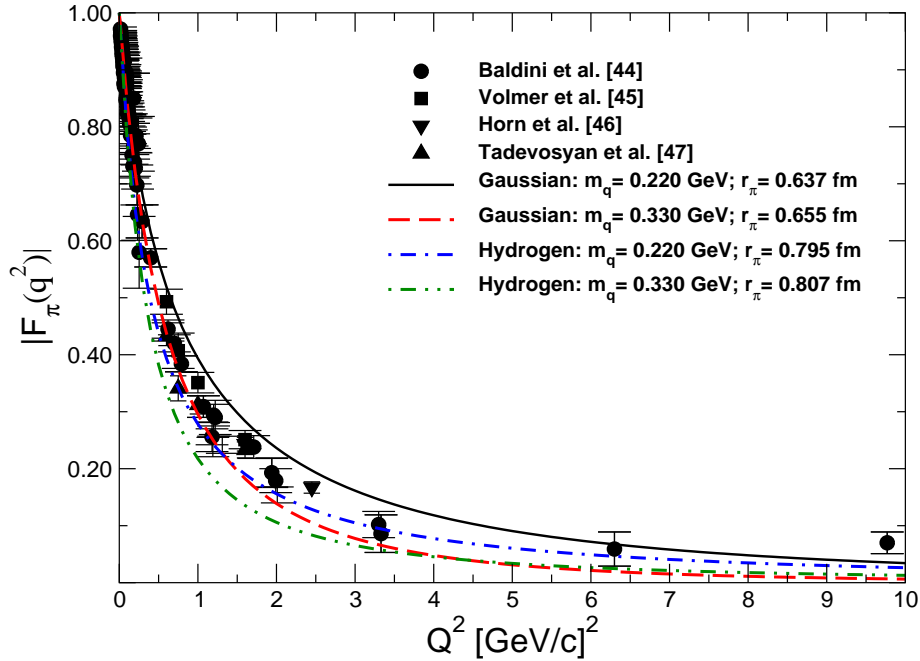}
\caption{The space-like elastic form factor, $F_\pi(Q^2)$, for two different constituent quark masses $m_q=m_u= m_d=M$ and $f_\pi = 92.4$~MeV; see 
Eqs.~\eqref{gauss} and \eqref{h2o} for the two model wave functions and Table~\ref{table1} for the corresponding length scale parameter, $r_\mathrm{nr}$. 
Data are from Ref.~\cite{Baldini:2000sh,Volmer:2000ek,Horn:2006tm,Tadevosyan:2007yd}.}
\label{fig4}
\end{figure}

\begin{figure}[t]
\vspace*{-4mm}
\centering
\includegraphics*[scale=0.5]{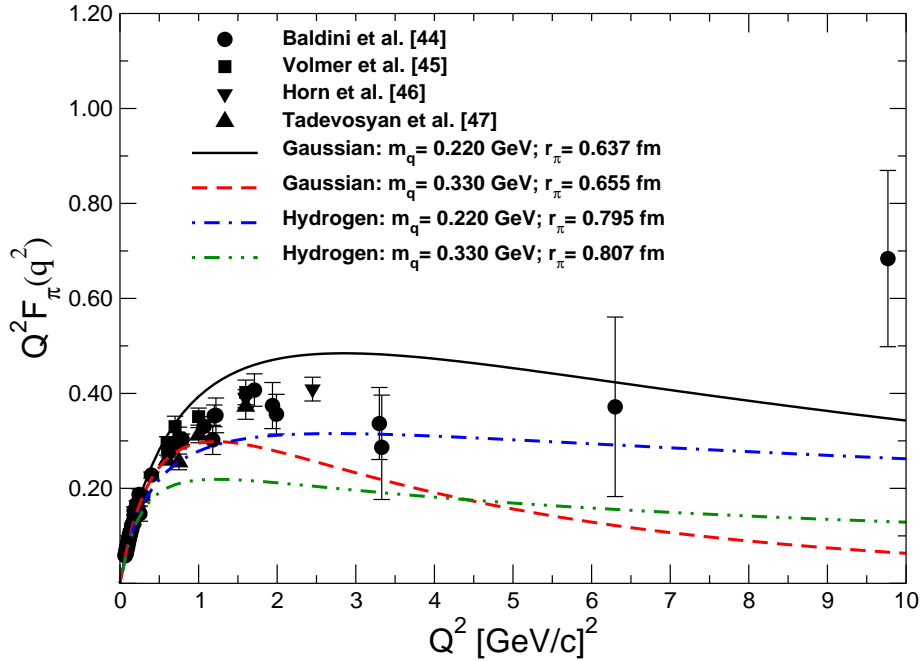}
\caption{The momentum-squared weighted elastic form factor, $Q^2 F_\pi(Q^2)$, for two different constituent quark masses $m_q=m_u= m_d=M$ and $f_\pi = 92.4$~MeV; 
see Eqs.~\eqref{gauss} and \eqref{h2o} for  the two model wave functions and Table~\ref{table1} for the corresponding length scale parameter, $r_\mathrm{nr}$.
Data are from Ref.~\cite{Baldini:2000sh,Volmer:2000ek,Horn:2006tm,Tadevosyan:2007yd}.}
\label{fig5}
\end{figure}

\begin{figure}[t]
\centering
\includegraphics*[scale=0.5]{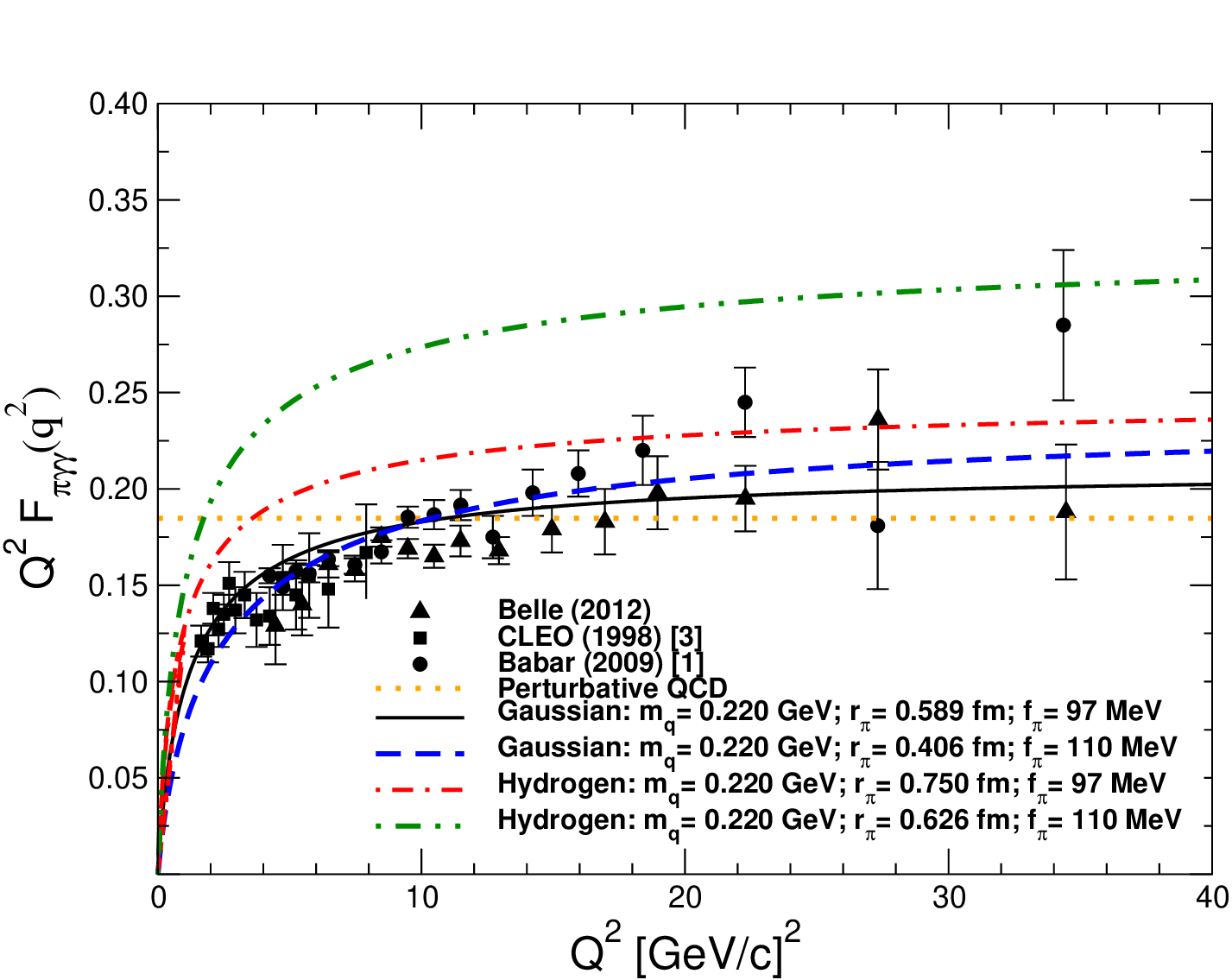}
\caption{The momentum-squared weighted transition form factor, $Q^2 F_{\gamma \pi}(Q^2)$,  for $m_q=m_u=m_d=M=220$~MeV and two values of $f_\pi > 92.4$ MeV.
See Table~\ref{table2} for the corresponding length scale parameter, $r_\mathrm{nr}$. Data are from Refs.~\cite{Aubert:2009mc,Gronberg:1997fj,Uehara:2012ag}. }
\label{fig7}
\end{figure}

\begin{figure}[h!]
\centering
\includegraphics*[scale=0.5]{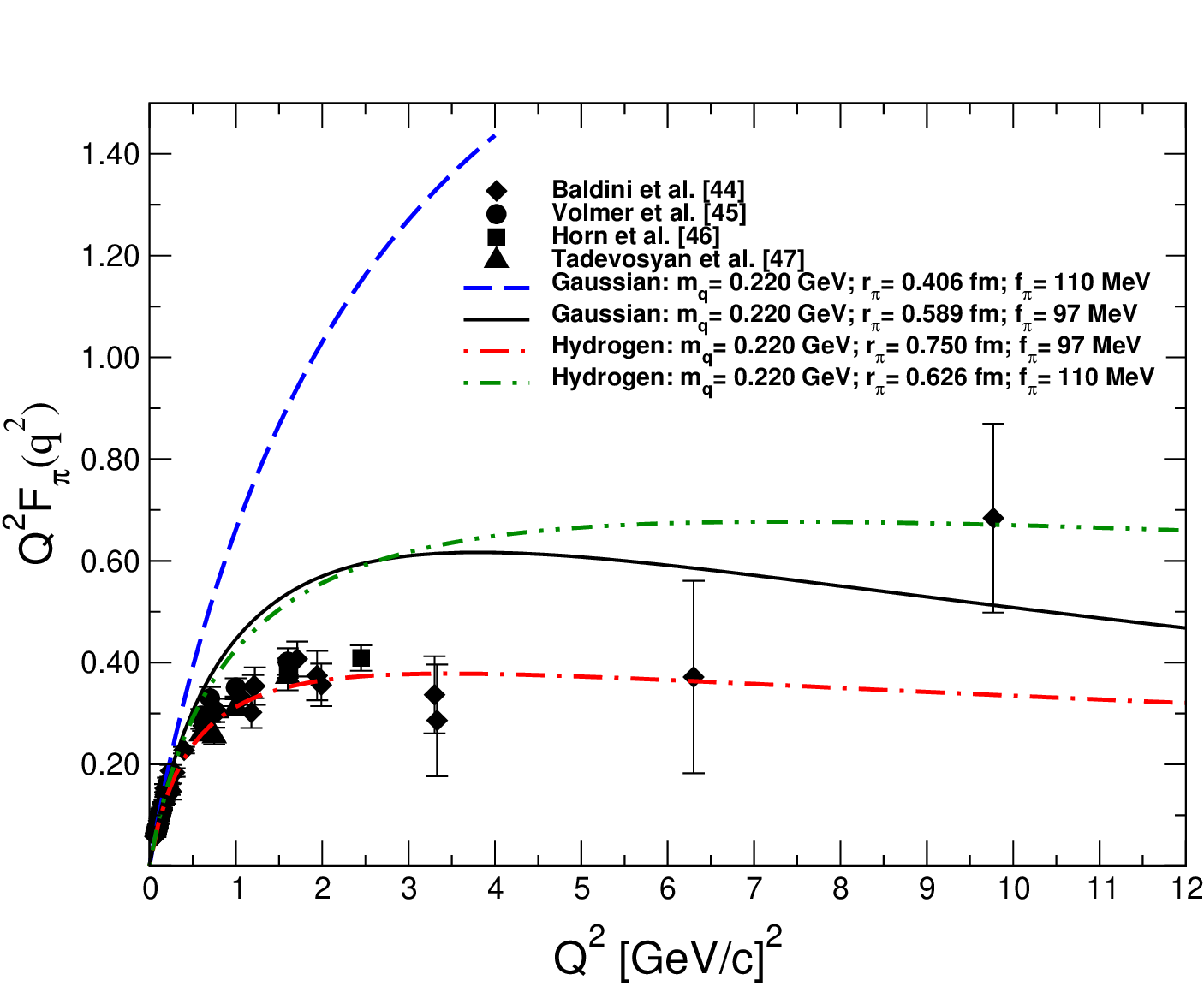}
\caption{The momentum-squared weighted elastic form factor, $Q^2 F_\pi(Q^2)$,  for $m_q=m_u=m_d=M=220$~MeV and two values of $f_\pi > 92.4$ MeV.
See Table~\ref{table2} for the corresponding length scale parameter, $r_\mathrm{nr}$. Data are from Ref.~\cite{Baldini:2000sh,Volmer:2000ek,Horn:2006tm,Tadevosyan:2007yd}. }
\label{fig8}
\end{figure}

\section{BaBar data, the pion's length scale and the impact on the charge radius}
\label{four}

Having fixed mass and length scale parameters of the form factors derived in Section~\ref{two}, we now allow for variation of the latter scale, 
$r_\mathrm{nr}$, while keeping $m_{u,d}=220$~MeV fixed. Table~\ref{table2} illustrates that when $r_\mathrm{nr}$ decreases
so does the rms charge radius, $\langle r_\pi^2 \rangle^{1/2}$, whereas the weak decay constant, $f_\pi$, increases in agreement with relation~\eqref{Tarrach}. 
This is observed independently of the wave function model employed. Bearing this on mind, we modify $r_\mathrm{nr}$, in other words the pion width 
in momentum space, and observe the impact on the form factors discussed in Section~\ref{three}.

The results for the transition and elastic form factors weighed by $Q^2$ are shown for two different charge radii and corresponding decay constants (see Table~\ref{table2})
in Figs.~\ref{fig7} and \ref{fig8} and for comparison, we plot the form factors for both models. The comparison within Fig.~\ref{fig7} and between Figs.~\ref{fig3} and \ref{fig7}
highlights that the model calculation of $Q^2 F_{\gamma \pi^0}(Q^2)$ coincide increasingly better with BaBar data for smaller charge radii. In particular, the hydrogen type 
bound-state function for which $f_\pi=97$~MeV, viz. about 5\% larger than in Table~\ref{table1}, yields a transition form factor in reasonable agreement with BaBar data 
for $Q^2 >15~[\mathrm{GeV}/c]^2.$ Nonetheless, below $Q^2 \simeq 15~[\mathrm{GeV}/c]^2$, this model produces a transition form factor much harder than experimentally 
evidenced. For the larger decay constant,  $f_\pi =110$~MeV, this is even more so the case. At any rate, the wave function model is not appropriate since it neither reproduces 
the extent CLEO data nor the BaBar data  which hints at a logarithmically increasing form factor at large $Q^2$.

The best result depicted in Fig.~\ref{fig7} is given by the dashed (blue) line for the Gaussian model and with $f_\pi = 110$~MeV from which we deduce
$\langle r_\pi^2 \rangle^{1/2} =  0.406$~[fm], {\em i.e.\/} a pion charge radius 40\% smaller than its experimental value~\cite{Beringer:1900zz}. Decreasing even more
$r_\mathrm{nr}$ will lead to a harder form factor $F_{\gamma \pi^0}(Q^2)$ and improve the agreement with BaBar data, however the corresponding weak decay constant and
charge radius turn out to take unrealistic values. Moreover, within the light-cone framework, a wave-function model congruent with the apparent logarithmic increase
of BaBar's $F_{\gamma \pi^0}(Q^2)$ data at large $Q^2$, e.g. the dashed (blue) line in Fig.~\ref{fig7}, is shown in Fig.~\ref{fig8} to fail in reasonably describing the elastic 
form factor,  $F_\pi(Q^2)$. The hydrogen-type wave functions yield reasonable rms charge radii and $F_\pi(Q^2)$ form factors yet, as noted before, the transition form factors
are poorly described in most of the $Q^2$ domain. Unsurprisingly, here too smaller charge radii are the results of harder form factors which are in disagreement with the 
current set of data~\cite{Baldini:2000sh,Volmer:2000ek,Horn:2006tm,Tadevosyan:2007yd} on the elastic pion form factor.

\section{Conclusive Remarks}
\label{five}

A series of recent articles \cite{Radyushkin:2009zg,Dorokhov:2010zz,Arriola:2010aq,Kroll:2010bf,Wu:2010zc,Pham:2011jq,Agaev:2010aq,Lichard:2010ap,Gorchtein:2011vf,
Roberts:2010rn,Agaev:2012tm,Stefanis:2011fv,Bakulev:2011rp,Bakulev:2012nh,Brodsky:2011yv,Brodsky:2011xx,Balakireva:2011wp} have dealt with the discrepancy between
BaBar data~\cite{Aubert:2009mc} on $\gamma^*\gamma\to\pi^0$ and a long-standing prediction of pQCD~\cite{Farrar:1979aw,Lepage:1980fj}. Attempts to reproduce the 
BaBar data can be generally identified within two broad classes:
\begin{description}
\item[$\bullet$ ]  alteration of the asymptotic form of the pion's distribution amplitude or wave function; \vspace*{2mm}
\item[$\bullet$ ]  dressing of the $\gamma \bar qq$ vertex with phenomenological interactions, such as vector dominance.
\end{description}

As has been discussed extensively in Ref.~\cite{Roberts:2010rn,Stefanis:2011fv,Bakulev:2011rp,Bakulev:2012nh,Brodsky:2011yv,Brodsky:2011xx} the modifications of the pion distribution 
amplitude, as proposed for instance in Refs.~\cite{Radyushkin:2009zg,Dorokhov:2010zz,Kroll:2010bf,Wu:2010zc,Pham:2011jq,Agaev:2010aq}, deviate drastically from its QCD asymptotic form. 
These  altered distributions, $\phi(x)$, are constant or at least non-vanishing for $x=0,1$ and characterize an essentially point-like pion. One comes to a similar conclusion within the
framework of a constituent quark model on the light cone or, for the matter, with any other quark model. A faithful reproduction of the BaBar data at large-$Q^2$ values leads
to a non-vanishing asymptotic pion wave function in momentum space\footnote{at least for practical purposes, as we use Gaussian and hydrogen wave function models
which eventually do vanish at large enough $k^2$ even for extreme values of $r_{\mathrm{nr}}$.}, which we showed to correspond to an excessively small pion charge radius. 
More generally, if the bound-state wave function does not asymptotically fall off as (at least) $k^{-2}$, the elastic form factor, $F_\pi(Q^2)$, becomes harder and may even turn 
out to take a constant value at large $Q^2$. This contradicts the bulk of existing data on the elastic pion form factor. We conclude that it is impossible, within a given consistent approach, 
to describe $f_\pi$, $\langle r_\pi^2 \rangle^{1/2}$,  $F_\pi(Q^2)$ and $F_{\gamma\pi} (Q^2)$ equally well. One may alter the pion bound-state function to account for the 
large-$Q^2$  BaBar data; this, however conflicts with other data and QCD-based studies that produce soft pions, viz. pion distribution amplitudes that vanish as $\sim (1-x)^2$ 
for $x\sim 1$. We also note that our pion transition form factor, the solid line in Fig.~\ref{fig7}, is in very good agreement with predictions by Agaev  {\em et al.\/}~\cite{Agaev:2012tm},
Bakulev {\em et al.\/}~\cite{Bakulev:2012nh} and Brodsky {\em et al.\/}~\cite{Brodsky:2011yv,Brodsky:2011xx}.

With respect to the photon vertex modification~\cite{Arriola:2010aq,Lichard:2010ap,Gorchtein:2011vf}, whilst resonances dominate the electromagnetic form factor of the
process $\gamma^* \to \pi^+\pi^-$ in the time-like region, this is not necessarily true in the deep space-like domain. It is reasonable to assume the imaginary part of the form factor 
vanishes at large time-like momentum squared, $s=q^2\to \infty$, and has a unitary cut on the real axis starting at the threshold $s=4m_\pi^2$. In this case, the real part is obtained 
from an unsubtracted dispersion relation,
\begin{equation}
 F_\pi (s) = \frac{1}{\pi} \int_{4m_\pi^2}^\infty \! \! ds' \ \frac{\mathrm{Im} \, F_\pi (s') }{s-s'} \ .
\label{disprel}
\end{equation}
The imaginary part receives contributions from intermediate resonances that dress the quark-photon vertex. Most of these intermediate states, $\gamma^*\to X \to \pi^+\pi^-$, lie below
$s=10$~GeV$^2$, the most prominent of which being the $\rho^0(770)$. However, the integration in Eq.~\eqref{disprel} is over infinite values of $s'$ and thus, in averaging out over all 
possible intermediate states, these resonances are unlikely to have a big impact on the real part of $F_\pi (s)$. 

Indeed, analyticity and unitarity suggest that higher resonances have little impact on the space-like elastic form factor even at high momentum transfers, which was shown with a 
phenomenological light-front constituent quark model that incorporates photon dressing \cite{deMelo:2003uk,deMelo:2005cy}. This model describes the space-like region up to 
$-10$~[GeV/$c]^2$, while in time-like region results are  reasonable up to $10~[\mathrm{GeV}/c]^2$. The approach was developed to calculate the elastic pion form factor in the 
space- and time-like regions starting from the covariant Mandelstam formula. A vector meson dominance model, built microscopically from the resolvent of a light front mass-squared 
operator in the spin-1 channel, was used for the quark-photon vertex dressing. This resolvent is obtained from the vector meson resonance wave functions.  The wave functions of 
both the pion and vector meson resonances are eigenstates of the relativistic constituent quark squared-mass operator \cite{Frederico:2002im,Frederico:2002vs}, which accounts 
for both confinement through a harmonic oscillator potential and $\pi-\rho$ splitting by means of a Dirac-delta interaction in the pseudoscalar channel. The coupling of the vector  
resonances to the photon and the decay vertex of the vector meson to the quark-antiquark pair are calculated within the model. The vector decay constants and the overlap between 
the vector-meson resonances and pion light-front wave functions decrease fast with the radial excitation. This is because the increasing number of nodes in the excited-state 
wave functions of vector mesons naturally damps both the decay constant as well as the overlap with the pion, strongly suppressing the contribution of the higher resonances 
to the space-like pion form factor. This behavior can be appreciated in Fig.~9 of Ref.~\cite{deMelo:2005cy} where it is seen that the higher resonances have no influence on 
the functional behavior of the elastic form factor.

If an analogous model is applied to the $\gamma\gamma^*\to \pi^0$ transition form factor, one expects the higher resonances to be suppressed as elucidated by the general discussion 
based on the dispersion relations in Eq.~(\ref{disprel}). There is no obvious sign that any amount of dressing the quark-photon vertex with higher resonances can significantly alter the 
asymptotic  $q^{-2}$ decrease of the form factor, whether in the region $Q^2=20-40~[\mathrm{GeV}/c]^2$ or below. Of course, this issue merits further investigation, which we postpone 
to future scrutiny. 

The results of our study suggest the large-$Q^2$ BaBar data are inconsistent with other experimental data on form factors, in particular the Belle data, and static properties of the pion. 
These are, in turn, in agreement with long-standing numerical results of pQCD and soft QCD. Moreover, it is intriguing that the $\gamma^* \to\eta \gamma$ and  $\gamma^*\to\eta' \gamma$ 
transition form factors also measured by the BaBar Collaboration at $Q^2=112~[\mathrm{GeV}/c]^2$ \cite{Aubert:2006cy} are in full agreement with the CLEO  results~\cite{Gronberg:1997fj}
and theoretical results, see for instance the green band in Fig.~2 of Ref.~\cite{Bakulev:2012nh}.

\begin{acknowledgements}
This work was supported by FAPESP grants nos.~2009/00069-5, 2009/53351-0, 2009/51296-1 and 2010/05772-3 and CNPq grants no.~306395/2009-6 and 305131/2011-7. 
Constructive and enjoyable discussions with Adnan Bashir, Craig Roberts and Peter Tandy were strongly appreciated. We acknowledge pertinent comments on the manuscript
by Nicos Stefanis.
\end{acknowledgements}

\bibliographystyle{h-physrev}
\bibliography{biblioSPIRES}

\end{document}